\documentclass{pjab}
\usepackage[T1]{fontenc}
\usepackage{lmodern}
\usepackage{textcomp}
\usepackage{amsmath}
\usepackage[dvipdfmx]{graphicx}
\usepackage{color}
\usepackage{array}
\usepackage[superscript]{cite}

\begin{document}
\pjabcategory{Review}
\title[Weak lensing mass map peaks]
      {Peaks in weak lensing mass maps for cluster astrophysics and cosmology}
\authorlist{%
 \Cauthorentry{Masamune Oguri}{aff1,aff2}
 \authorentry{Satoshi Miyazaki}{aff3,aff4}
}
\affiliate[aff1]{Center for Frontier Science, Chiba University, 1-33 Yayoi-cho, Inage-ku, Chiba 263-8522, Japan}
\affiliate[aff2]{Department of Physics, Graduate School of Science, Chiba University, 1-33 Yayoi-Cho, Inage-Ku, Chiba 263-8522, Japan}
\affiliate[aff3]{Subaru Telescope, National Astronomical Observatory of Japan, 96720, Hilo, HI, USA}
\affiliate[aff4]{SOKENDAI, Graduate University for Advanced Studies,  181-8588, Osawa, Mitaka, Tokyo, Japan}
\Correspondence{M. Oguri, Center for Frontier Science, Chiba University, 1-33 Yayoi-cho, Inage-ku, Chiba 263-8522, Japan, masamune.oguri@chiba-u.jp}
\abstract{
 Clusters of galaxies can be identified from peaks in weak lensing aperture mass maps constructed from weak lensing shear catalogs. Such purely gravitational cluster selection considerably differs from traditional cluster selections based on baryonic properties of clusters. In this review, we present the basics and applications of weak lensing shear-selected cluster samples. Detailed studies of baryonic properties of shear-selected clusters shed new light on cluster astrophysics. The purely gravitational selection suggests that the selection function can be quantified more easily and robustly, which is crucial for deriving accurate cosmological constraints from the abundance of shear-selected clusters. The recent advance of shear-selected cluster studies is driven by the Subaru Hyper Suprime-Cam survey, in which more than 300 shear-selected clusters with the signal-to-noise ratio greater than 5 are identified. It is argued that various systematic effects in the cosmological analysis can be mitigated by carefully choosing the set-up of the analysis, including the choice of the kernel functions and the source galaxy sample.
}
\maketitle

\section{Introduction}

The standard $\Lambda$-dominated Cold Dark Matter ($\Lambda$CDM) cosmological model requires the presence of two unknown components called dark matter and dark energy. The nature of dark matter and dark energy is one of the central problems in modern cosmology. It is expected that a clue to their nature is obtained by detailed studies of the structure of the University, because the information on both the expansion history of the Universe as well as the nature of dark matter is imprinted in the matter distribution of the Universe\cite{2022JHEAp..34...49A}. Since the matter distribution is dominated by dark matter that cannot directly be observed, we need to resort to indirect methods to study the distribution of dark matter.

Weak gravitational lensing, which takes advantage of small distortions (``shears'') of distant galaxies due to the deflection of light rays by intervening gravitational fields, provides a powerful means of studying the distribution of dark matter\cite{2001PhR...340..291B}. The gravitational lensing effect can be robustly predicted by General Relativity, and allows us to directly map the total mass distribution including dark matter.  

Clusters of galaxies provide another means of studying the matter distribution in the Universe\cite{2011ARA&A..49..409A}. Clusters are the most massive gravitationally bound objects in the Universe, and their abundance as well as the internal structure are mainly determined by the gravitational dynamics of dark matter. Clusters of galaxies can be easily and securely identified in several different observations, including optical imaging and spectroscopy to identify member galaxies and observations of X-ray and the Sunyaev-Zel'dovich effect originating from hot gas in clusters of galaxies\cite{2005RvMP...77..207V}. 

Clusters of galaxies can also be identified directly from weak gravitational lensing data\cite{1993ApJ...404..441K,1996MNRAS.283..837S}. The small distortions of galaxies can be inverted to the projected mass distribution (``convergence''), which is referred to as a mass map, as both convergence and shear are calculated by second derivatives of the lens potential. Clusters of galaxies are identified as high signal-to-noise ratio peaks in weak lensing mass maps. Such identification of clusters represents a purely gravitational selection and is insensitive to complicated baryon physics, which is a clear advantage over the selections of clusters in optical, X-ray, and the Sunyaev-Zel'dovich effect.

In this paper, we review clusters of galaxies selected from mass map peaks, which are sometimes referred to as shear-selected clusters. As detailed below, constructing a large sample of shear-selected clusters require deep {\em and} wide imaging surveys, which have been enabled only recently. We describe how such shear-selected clusters advance our understanding of clusters of galaxies, and also how those clusters can be used to place tight constraints on cosmological parameters. We note that, for the purpose of constraining cosmological parameters, mass map peaks alone without the association with clusters of galaxies are sometimes sufficient as observables that can be predicted from cosmological models. In this review, we rather focus on mass map peaks as proxies for galaxy clusters.

\section{Basics of gravitational lensing}

\subsection{Lens equation}

Under the geometric optics approximation, the propagation of light is determined by the geodesic equation
\begin{equation}
  \frac{d^2x^\mu}{d\lambda^2}
  +\Gamma^{\mu}{}_{\alpha\beta}\frac{dx^\alpha}{d\lambda}\frac{dx^\beta}{d\lambda} =0,
  \label{eq:geodesic_def_geoeq}
\end{equation}
where $\lambda$ is an affine parameter and $\Gamma^{\mu}{}_{\alpha\beta}$ are the Christoffel symbols. It is often useful to rewrite the geodesic equation as
\begin{equation}
  \frac{d}{d\lambda}\left(g_{\mu\nu}\frac{dx^\nu}{d\lambda}\right)
  -\frac{1}{2}g_{\alpha\beta,\mu}\frac{dx^\alpha}{d\lambda}\frac{dx^\beta}{d\lambda}=0,
  \label{eq:geodesic_def_geoeq_ano}
\end{equation}
where $g_{\alpha\beta}$ is a metric tensor and a comma represents a partial derivative.

In an expanding Universe with inhomogeneous density fluctuations, the line element is given by
\begin{align}
ds^2 &=-\left(1+\frac{2\Phi}{c^2}\right)c^2dt^2+a^2
\left(1-\frac{2\Psi}{c^2}\right)\nonumber\\
&\quad\times\left[d\chi^2+f_K^2(\chi)\omega_{ab}dx^adx^b\right],
\label{eq:lenseq_metric_newtonian}
\end{align}
where $a$ is the scale factor, $\Phi$ is the gravitational potential, $\Psi$ is the curvature perturbation, 
\begin{equation}
f_K(\chi) =
\begin{cases}
    {\displaystyle \frac{1}{\sqrt{K}}}\sin\left(\sqrt{K}\chi\right), & (K>0)\\
    \chi, & (K=0)\\
    {\displaystyle \frac{1}{\sqrt{-K}}}\sinh\left(\sqrt{-K}\chi\right), & (K<0)
\end{cases}
\label{eq:geodesic_def_fk}
\end{equation}
where $K$ is the spatial curvature of the Universe, and
\begin{equation}
\omega_{ab}dx^adx^b = d\theta^2+\sin^2\theta d\phi^2.
\label{eq:lenseq_def_metric_omega}
\end{equation}

By plugging the metric tensor defined by Eq.~\eqref{eq:lenseq_metric_newtonian} in Eq.~\eqref{eq:geodesic_def_geoeq_ano} and computing the angular part ($\mu=a$) of the equation, we obtain
\begin{equation}
  \frac{d}{d\chi}\left[f_K^2(\chi)\frac{dx^a}{d\chi}\right]
  +\frac{1}{c^2}\omega^{ab}\left(\Phi_{,b}+\Psi_{,b}\right)=0.
  \label{eq:lenseq_lenseq_diff_pre}
\end{equation}
Since we have $\Phi=\Psi$ from the Einstein equations without anisotropic stress, in what follows we always set $\Phi=\Psi$ and rewrite Eq.~\eqref{eq:lenseq_lenseq_diff_pre} as
\begin{equation}
   \frac{d}{d\chi}\left[f_K^2(\chi)\frac{dx^a}{d\chi}\right]
  +\frac{2}{c^2}\omega^{ab}\Phi_{,b}=0.
  \label{eq:lenseq_lenseq_diff}
\end{equation}
Eq.~\eqref{eq:lenseq_lenseq_diff_pre} can also be obtained by the Fermat's principle, considering the variation of the conformal time $d\eta=dt/a$ integrated along the line-of-sight and using the Euler-Lagrange equation.

The lens equation for a source at $\chi=\chi_{\mathrm{s}}$ is obtained by integrating Eq.~\eqref{eq:lenseq_lenseq_diff} twice 
\begin{equation}
  x^a(\chi_{\mathrm{s}})-x^a(0)=
  -\frac{2}{c^2}\int_0^{\chi_{\mathrm{s}}}d\chi
  \frac{f_K(\chi_{\mathrm{s}}-\chi)}{f_K(\chi)f_K(\chi_{\mathrm{s}})}
  \omega^{ab}\Phi_{,b}.
  \label{eq:lenseq_lenseq1}
\end{equation}
The right hand side of Eq.~\eqref{eq:lenseq_lenseq1} represents the deflection angle. In cosmological gravitational lensing the deflection angle is always very small compared with a radian, and thus a locally flat sky coordinate is often adopted for the analysis. In addition, in the weak gravitational lensing analysis, the Born approximation, in which the integration in Eq.~\eqref{eq:lenseq_lenseq1} is evaluated along a straight line, is commonly adopted. With these approximations, Eq.~\eqref{eq:lenseq_lenseq1} is recast into
\begin{equation}
\boldsymbol{\beta}
=\boldsymbol{\theta}-\nabla_{\boldsymbol{\theta}}\psi,
\label{eq:lenseq_lens_equation_4}
\end{equation}
where $\boldsymbol{\beta}$ ($=x^a(\chi_{\mathrm{s}})$) is the source position on the sky, $\boldsymbol{\theta}$ ($=x^a(0)$) is the image position on the sky, $\nabla_{\boldsymbol{\theta}}$ is the gradient vector with respect to the image position $\boldsymbol{\theta}$, and $\psi$ is the lens potential defined by
\begin{equation}
\psi(\boldsymbol{\theta})=\frac{2}{c^2}\int_0^{\chi_{\mathrm{s}}}d\chi
\frac{f_K(\chi_{\mathrm{s}}-\chi)}{f_K(\chi)f_K(\chi_{\mathrm{s}})}
\Phi(\chi, \boldsymbol{\theta}).
\label{eq:lenseq_lens_potential_1}
\end{equation}
Eq.~\eqref{eq:lenseq_lens_potential_1} indicates that the lens potential is given by the gravitational potential integrated along the line-of-sight with a weight.

\subsection{Convergence and shear}
Eq.~\eqref{eq:lenseq_lens_equation_4} suggests that the shape of the image $\delta\boldsymbol{\theta}$ is related with that of the source $\delta\boldsymbol{\beta}$ as 
\begin{equation}
  \delta\boldsymbol{\beta}=A(\boldsymbol{\theta})\delta\boldsymbol{\theta},
  \label{eq:lensprop_image_deform}
\end{equation}
where the Jacobi matrix $A(\boldsymbol{\theta})$ is expressed by second derivatives of the lens potential in the image plane $\boldsymbol{\theta}=(\theta_1,\,\theta_2)$ and is given by
\begin{equation}
  A(\boldsymbol{\theta})
    =
\begin{pmatrix}
1-\psi_{,\theta_1\theta_1} & -\psi_{,\theta_1\theta_2} \\
-\psi_{,\theta_1\theta_2} & 1-\psi_{,\theta_2\theta_2} \\
\end{pmatrix},
\label{eq:lensprop_jacobi_matrix2}
\end{equation}
which is rewritten as
\begin{equation}
  A(\boldsymbol{\theta})=
\begin{pmatrix}
1-\kappa-\gamma_1 & -\gamma_2 \\
-\gamma_2 & 1-\kappa+\gamma_1 \\
\end{pmatrix},
\label{eq:lensprop_jacobi_matrix3}
\end{equation}
where $\kappa=(\psi_{,\theta_1\theta_1}+\psi_{,\theta_2\theta_2})/2$ is convergence and $\gamma_1=(\psi_{,\theta_1\theta_1}-\psi_{,\theta_2\theta_2})/2$ and $\gamma_2=\psi_{,\theta_1\theta_2}$ are shear. From Eq.~\eqref{eq:lenseq_lens_potential_1}, it is shown that convergence can be connected with the matter density fluctuation $\delta_{\mathrm{m}}=(\rho_{\mathrm{m}}-\bar{\rho}_{\mathrm{m}})/\bar{\rho}_{\mathrm{m}}$ as
\begin{align}
\kappa(\boldsymbol{\theta}) &=\frac{4\pi G }{c^2}\int_0^{\chi_{\mathrm{s}}}d\chi
\frac{f_K(\chi_{\mathrm{s}}-\chi)f_K(\chi)}{f_K(\chi_{\mathrm{s}})}\nonumber\\
&\quad\times\bar{\rho}_{\mathrm{m}}a^2\delta_{\mathrm{m}}(\chi, \boldsymbol{\theta}),
\label{eq:lensprop_kappa_born_calc2}
\end{align}
using the Poisson's equation for the gravitational potential. This indicates that convergence is essentially the matter density field projected along the line-of-sight with a weight.

\subsection{Mass map reconstruction}

In weak gravitational lensing, the average shear in some region on the sky is estimated by averaging shapes of galaxies in the region, assuming that intrinsic orientations of galaxies are random. Once the complex shear field $\gamma(\boldsymbol{\theta})=\gamma_1(\boldsymbol{\theta})+i\gamma_2(\boldsymbol{\theta})$ is constructed from average shapes of galaxies, we can convert $\gamma(\boldsymbol{\theta})$ to $\kappa(\boldsymbol{\theta})$ as
\begin{equation}
  \kappa(\boldsymbol{\theta})=\frac{1}{\pi}\int d\boldsymbol{\theta}' \gamma(\boldsymbol{\theta}')
  D^*(\boldsymbol{\theta}-\boldsymbol{\theta}'),
  \label{eq:wl_kaiser_squires}
\end{equation}
\begin{equation}
  D(\boldsymbol{\theta})= \frac{\theta_2^2-\theta_1^2-2i\theta_1\theta_2}{\left|\boldsymbol{\theta}\right|^4},
\end{equation}
because both convergence and shear are described by second derivatives of the same lens potential.

However, in practice, the reconstructed convergence map is heavily affected by the shot noise arising from the fact that we use a discrete galaxy sample to estimate the shear field. In order to suppress the shot noise, we first smooth the shear field by the Gaussian kernel
\begin{equation}
  W_{\mathrm{s}}(\boldsymbol{\theta})=\frac{1}{\pi\sigma_{\mathrm{s}}^2}\exp\left(-\frac{\left|\boldsymbol{\theta}\right|^2}{\sigma_{\mathrm{s}}^2}\right),
  \label{eq:wl_kernel_ws}
\end{equation}
to obtain the smoothed shear field as
\begin{equation}
  \gamma_{\mathrm{s}}(\boldsymbol{\theta})=\int d\boldsymbol{\theta}' \gamma(\boldsymbol{\theta}') W_{\mathrm{s}}(\boldsymbol{\theta}-\boldsymbol{\theta}'),
 \label{eq:wl_gamma_s}
\end{equation}
and use the smoothed shear field to reconstruct the convergence field. The resulting convergence field is a smoothed convergence field with the same kernel function, Eq.~\eqref{eq:wl_kernel_ws}. This method is called the Kaiser-Squires method\cite{1993ApJ...404..441K}.

It is known that Eqs.~\eqref{eq:wl_kaiser_squires} and \eqref{eq:wl_gamma_s} are also described using the tangential shear\cite{1996MNRAS.283..837}
\begin{equation}
\gamma_+(\boldsymbol{\theta}';\,\boldsymbol{\theta}) =
-\mathrm{Re}\left[\gamma(\boldsymbol{\theta}')\,  e^{-2i\varphi_{\boldsymbol{\theta}'-\boldsymbol{\theta}}}\right],
\end{equation}
where $\varphi_{\boldsymbol{\theta}'-\boldsymbol{\theta}}$ is the polar angle of $\boldsymbol{\theta}'-\boldsymbol{\theta}$. The smoothed convergence field $\kappa_{\mathrm{s}}(\boldsymbol{\theta})$ is given by
\begin{align}
  \kappa_{\mathrm{s}}(\boldsymbol{\theta})&=\int d\boldsymbol{\theta}' \kappa(\boldsymbol{\theta}')W_{\mathrm{s}}(\boldsymbol{\theta}-\boldsymbol{\theta}')\nonumber\\
  &=\int d\boldsymbol{\theta}' \gamma_+(\boldsymbol{\theta}';\,\boldsymbol{\theta}) Q_{\mathrm{s}}(|\boldsymbol{\theta}-\boldsymbol{\theta}'|),
  \label{eq:def_kappa_s}
\end{align}
where
\begin{equation}
  Q_{\mathrm{s}}(\theta)=\frac{1}{\pi\theta^2}\left[1-\left(1+\frac{\theta^2}{\sigma_{\mathrm{s}}^2}\right)\exp\left(-\frac{\theta^2}{\sigma_{\mathrm{s}}^2}\right)\right].
  \label{eq:qs_gauss}
\end{equation}

\subsection{Aperture mass map} We can generalize Eq.~\eqref{eq:def_kappa_s} to consider the aperture mass map\cite{1996MNRAS.283..837S,1998MNRAS.296..873S}, which is the convergence field convolved with a kernel $U(\theta)$. Specifically, the aperture mass map $M_{\mathrm{ap}}(\boldsymbol{\theta})$ is defined as
\begin{equation}
 M_{\mathrm{ap}}(\boldsymbol{\theta})=\int d\boldsymbol{\theta}'
 \kappa(\boldsymbol{\theta}')U(|\boldsymbol{\theta}-\boldsymbol{\theta}'|).
 \label{eq:map_def}
\end{equation}
Assuming that the filter $U(\theta)$ is a compensated filter
\begin{equation}
  \int_0^\infty d\theta\,\theta\,U(\theta)=0,
  \label{eq:u_comp}
\end{equation}
Eq.~\eqref{eq:map_def} can also be described using the tangential shear as
\begin{equation}
 M_{\mathrm{ap}}(\boldsymbol{\theta})=\int d\boldsymbol{\theta}' \gamma_+(\boldsymbol{\theta}';\,\boldsymbol{\theta}) Q(|\boldsymbol{\theta}-\boldsymbol{\theta}'|),
\label{eq:map_map_q}
\end{equation}
where $Q(\theta)$ is related with $U(\theta)$ as
\begin{equation}
  Q(\theta)=\frac{2}{\theta^2}\int_0^\theta d\theta'\theta'U(\theta')-U(\theta).
  \label{eq:from_u_to_q}
\end{equation}
Inversely, the filter $U(\theta)$ can be obtained from $Q(\theta)$ as
\begin{equation}
U(\theta)=\int_\theta^\infty d\theta'\frac{2}{\theta'}Q(\theta')-Q(\theta).
\end{equation}
As will be discussed later, by carefully choosing the kernel functions $U(\theta)$ and $Q(\theta)$ we can mitigate various systematic effects.

We caution that the locally flat sky is assumed for these calculations. The flay sky approximation is valid when the area of the mass map is not large, less than hundreds of square degrees. The extension to the curved sky is possible by using e.g., spin-weighted spherical harmonics \cite{2005PhRvD..72b3516C}.

\section{Searching for clusters of galaxies from mass map peaks}

\subsection{Construction of the mass map}

Since clusters of galaxies are quite efficient in producing weak lensing signals (see e.g., \cite{2020A&ARv..28....7U} for a review), they appear as high signal-to-noise ratio peaks in weak lensing aperture mass maps, depending on the filter\cite{1999A&A...345....1R,2000MNRAS.318..321K,2000ApJ...530L...1J,2004MNRAS.350..893H,2005ApJ...635...60T}. In practice, we have to reconstruct the aperture mass map from a discrete source galaxy sample. An estimator of Eq.~\eqref{eq:map_map_q} is given by
\begin{equation}
 \hat{M}_{\mathrm{ap}}(\boldsymbol{\theta})=\frac{1}{\bar{n}}\sum_j \gamma_+(\boldsymbol{\theta}_j;\boldsymbol{\theta}) Q(|\boldsymbol{\theta}-\boldsymbol{\theta}_j|),
\label{eq:map_est1}
\end{equation}
where $i$ runs over galaxies in the source galaxy sample and $\bar{n}$ is the average surface number density of source galaxies.

Since the precision of shape measurements of individual galaxies varies depending on sizes and magnitudes of galaxies, a weight is often introduced to optimize the weak lensing signal. In presence of the weight $w_j$ for the $j$-th galaxy, Eq.~\eqref{eq:map_est1} is modified as
\begin{equation}
 \hat{M}_{\mathrm{ap}}(\boldsymbol{\theta})=\frac{1}{\bar{w}\bar{n}}\sum_j w_j\gamma_+(\boldsymbol{\theta}_j;\boldsymbol{\theta}) Q(|\boldsymbol{\theta}-\boldsymbol{\theta}_j|),
\label{eq:map_est2}
\end{equation}
where $\bar{w}$ denote the average weight value.

\subsection{Signal-to-noise ratio}\label{sec:sn_ratio}

The important source of the noise in the aperture mass maps is intrinsic shapes of galaxies. By averaging intrinsic shapes of $N$ galaxies, we can reduce the noise of the estimation of the average shear by $1/\sqrt{N}$, suggesting that the number density of galaxies used for weak lensing determines the quality of the resulting mass map. Due to the consequence of the central limit theorem, the intrinsic shape noise translates into a random Gaussian noise of the weak lensing mass map, whose statistical property is well studied\cite{1986ApJ...304...15B}, to a good approximation\cite{2000MNRAS.313..524V}. The significance of the aperture mass map peak at $\boldsymbol{\theta}_{\mathrm{peak}}$ is usually quantified by the signal-to-noise ratio
\begin{equation}
\nu=\frac{M_\mathrm{ap}(\boldsymbol{\theta}_{\mathrm{peak}})}{\sigma},
\label{eq:def_nu}
\end{equation}
where $\sigma$ is the 1$\sigma$ noise of the aperture mass map. Suppose $\sigma_e=\sqrt{\sigma_{e_1}^2+\sigma_{e_2}^2}$ is the intrinsic shape noise of source galaxies, the variance of the tangential shear of each galaxy is dominated by the intrinsic shape noise and is given by
\begin{equation}
\langle \left\{\gamma_+(\boldsymbol{\theta}_j;\boldsymbol{\theta})\right\}^2\rangle = \frac{\sigma^2_e}{2}.
\end{equation}
Thus, from e.g., Eq~\eqref{eq:map_est2}, the 1$\sigma$ noise of the aperture mass map $\sigma$ is computed as
\begin{equation}
 \sigma^2=\frac{1}{2\bar{w}^2\bar{n}^2}\sum_j w_j^2\sigma_e^2 Q^2(|\boldsymbol{\theta}-\boldsymbol{\theta}_j|).
\label{eq:map_sigma}
\end{equation}
We note that the normalization of the aperture mass map $\hat{M}_{\mathrm{ap}}$ or the kernel function $Q(\theta)$ is not important as long as we use the signal-to-noise ratio $\nu$ defined by Eq.~\eqref{eq:def_nu} for selecting cluster candidates, as such normalization cancels out in calculating $\nu$.

The construction of the mass map and the calculation of the signal-to-noise ratio so far assume the uniform distribution of source galaxies on the sky. In practice, however, the number density of resolved galaxies on the sky is quite inhomogeneous due to e.g., the variation of observing conditions such as seeing sizes. In addition, some parts of the sky are masked due to e.g., the presence of bright foreground stars. As is obvious from Eqs.~\eqref{eq:map_est1} and \eqref{eq:map_est2}, the aperture mass map value at some point on the sky is given by shapes of surrounding source galaxies, and hence is significantly affected by the boundary effect.

One way to take account of such inhomogeneity and masking effects in measurements is to estimate the noise $\sigma$ as well as the average number density $\bar{n}$ and weight $\bar{w}$ locally as a function of the sky position $\boldsymbol{\theta}$. For instance, the local value of $\bar{w}\bar{n}$ at $\boldsymbol{\theta}$ can be estimated as
\begin{equation}
  \bar{w}\bar{n}\propto\sum_j w_jQ(|\boldsymbol{\theta}-\boldsymbol{\theta}_j|).
\label{eq:map_wn}
\end{equation}
Furthermore, a simple and powerful way to derive the noise $\sigma$ locally at each position on the sky is to randomize orientations of individual source galaxies and construct the aperture mass map, and repeat this many times to estimate $\sigma$ from the variance of randomized mass map values\cite{2002ApJ...580L..97M,2021PASJ...73..817O}. There are both advantages and disadvantages of such local estimations of the average density and noise\cite{2011ApJ...735..119S,2020PASJ...72...78H}, which depend also on the choice of the kernel functions that is discussed below. It is worth emphasizing that such inhomogeneity needs to be properly taken into account in cosmological inference, which is one of the main challenges for using mass map peaks or shear-selected clusters to constrain cosmological parameters.

\subsection{Choice of the kernel functions}

The choice of the kernel functions $U(\theta)$ and $Q(\theta)$ is important for an efficient search of clusters of galaxies from peaks in aperture mass maps. The signal-to-noise ratio is optimized by choosing the so-called matched filter, for which $Q(\theta)$ follows the the expected tangential shear profile $\gamma_+$ of the lensing cluster. Since it is known that the density profile of clusters of galaxies is well approximated by the Navarro-Frenk-White (NFW) density profile\cite{1997ApJ...490..493N}, both from theoretical predictions assuming $\Lambda$CDM as well as from weak lensing observations\cite{2011ApJ...738...41U,2013ApJ...769L..35O}, the functional form of $Q(\theta)$ that resembles the tangential shear profile predicted by the NFW profile has been considered\cite{1998MNRAS.296..873S,2003NewA....8..581P,2004A&A...420...75S,2005ApJ...624...59H,2005A&A...442..851M}.

In addition, it is important to reduce the contamination from the large-scale structure (LSS). The matter fluctuations in front of and behind halos of interest can sometimes enhance the signal-to-noise significantly, and such LSS is thought to be another important source of the noise when we construct a sample of massive clusters of galaxies from mass map peaks. In particular, aperture mass maps constructed with the simple Gaussian filter given by Eq.~\eqref{eq:qs_gauss} is known to be significantly affected by the LSS noise and hence is not ideal. In order to reduce the LSS noise, one has to choose kernel functions that suppress density fluctuations on scales much larger than typical sizes of clusters of galaxies\cite{2005A&A...442..851M,2012MNRAS.423.1711M}.

Weak lensing signals near the center of clusters are subject to various systematic uncertainties, including cluster member dilution effect\cite{2018PASJ...70...30M}, intrinsic alignments of cluster member galaxies\cite{2016MNRAS.463.3653K,2022ApJ...940...96Z}, the non-linearity of shear\cite{1995A&A...294..411S}, and the obscuration by cluster member galaxies\cite{2015MNRAS.449.1259S}. In this regard, choosing $Q(\theta)$ that is suppressed at small $\theta$ corresponding to the very central region of clusters is advantageous as signals are less affected by these systematics effects\cite{2021PASJ...73..817O}.

Finally, is it also useful to consider the kernel functions that are confined within a small finite radius, because it is less affected by the boundary effect\cite{1996MNRAS.283..837S}.

\subsection{Examples of aperture mass maps}

Here we present examples of aperture mass maps constructed with different kernel functions. In addition to the Gaussian smoothing kernel given by Eq.~\eqref{eq:qs_gauss}, we consider the power-law with the outer exponential-cutoff\cite{2012MNRAS.425.2287H}
\begin{equation}
Q(\theta) 
={(\theta/\theta_{\mathrm{in}})^n \over {\theta_{\mathrm{in}}^2(1+a \theta /\theta_{\mathrm{in}})^{(2+n)}}}
\exp\left(-{{\theta^2}\over {2 \theta_{\mathrm{out}}^2}}\right),
\label{eq:def_q_pex}
\end{equation}
with $(n,a)=(0,0.25)$ and $(1,0.7)$ being named as PEX0 and PEX1, respectively. PEX0 mimics the matched filter for the NFW density profile proposed\cite{2005A&A...442..851M}, whereas PEX1 approximates the kernel function that reduces the contribution of lensing signals from the inner part of clusters\cite{2005ApJ...624...59H}. For both PEX0 and PEX1, we adopt the inner scale radius of $\theta_{\mathrm{in}}=1.5'$ and the outer scale radius of $\theta_{\mathrm{out}}=7'$.

In addition, we consider the truncated isothermal kernel functions\cite{2021PASJ...73..817O}, which adopts the following functional form\cite{1996MNRAS.283..837S}
\begin{equation}
  U(\theta)=
  \left\{
  \begin{array}{ll}
    1 & (\theta \leq \nu_1\theta_R),\\
    U_1(\theta)
    & (\nu_1\theta_R \leq \theta \leq \nu_2\theta_R),\\
    U_2(\theta) 
    & (\nu_2\theta_R \leq \theta \leq \theta_R),\\
    0 & (\theta_R \leq \theta),
\end{array}
  \right.
  \label{eq:def_u_ti}
\end{equation}
\begin{equation}
U_1(\theta)=  \frac{1}{1-c} \left(
    \frac{\nu_1\theta_R}{\sqrt{(\theta-\nu_1\theta_R)^2+(\nu_1\theta_R)^2}}-c\right),
\end{equation}
\begin{equation}
U_2(\theta)=  \frac{b}{\theta_R^3}(\theta_R-\theta)^2(\theta-\alpha\theta_R),
\end{equation}
with $c$, $b$, $\alpha$ being determined from the requirement that $U(\theta)$ and its first derivative are continuous at $\theta=\nu_2\theta_R$ as well as the condition of the compensation filter given by Eq.~\eqref{eq:u_comp}. The parameters $\nu_1$, $\nu_2$, and $\theta_R$ are carefully chosen to maximize the signal-to-noise ratio. Specifically, for the TI05 setup $\nu_1=0.027$, $\nu_2=0.36$, and $\theta_R=18.5'$ are adopted, and for the TI20 setup, $\nu_1=0.121$, $\nu_2=0.36$, and $\theta_R=16.6'$ are adopted. We use Eq.~\eqref{eq:from_u_to_q} to convert $U(\theta)$ defined in Eq.~\eqref{eq:def_u_ti} to $Q(\theta)$.

\begin{figure*}[p]
\begin{center}
\includegraphics[width=0.8\hsize]{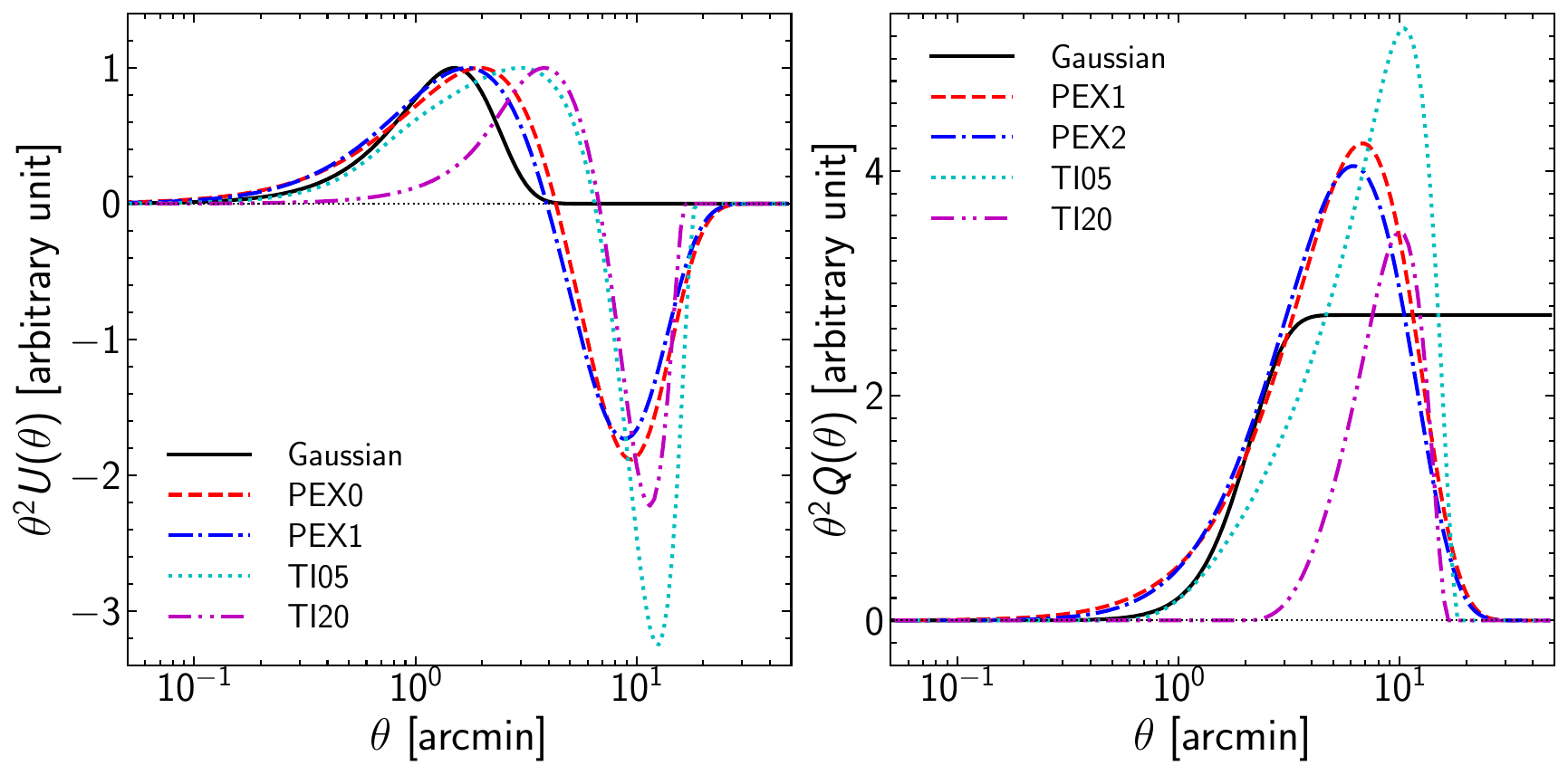}
\caption{The kernel functions $U(\theta)$ and $Q(\theta)$. See the text for the definitions of individual kernel functions.
\label{fig:filter}}
\end{center}
\end{figure*}
\begin{figure*}[p]
\begin{center}
\includegraphics[width=0.8\hsize]{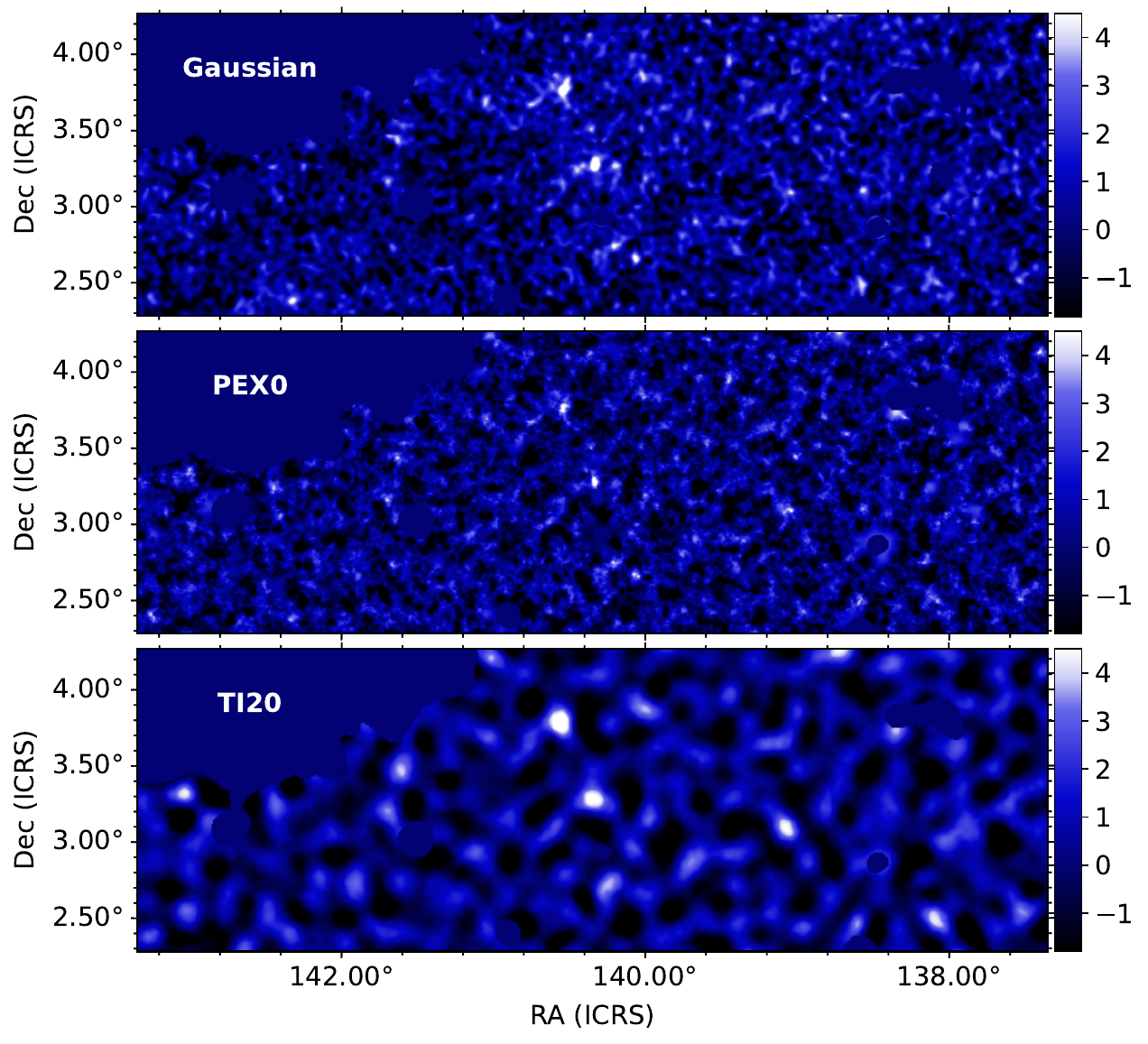}
\caption{Examples of aperture mass maps from the Hyper Suprime-Cam Subaru Strategic Program data\cite{2021PASJ...73..817O}. Aperture mass maps constructed with the Gaussian filter (Eq.~\eqref{eq:qs_gauss}), PEX0 (Eq.~\eqref{eq:def_q_pex}), and TI20 (Eq.~\eqref{eq:def_u_ti}) are compared. The color scale indicates the signal-to-noise ratio defined by Eq.~\eqref{eq:def_nu}.
\label{fig:maps}}
\end{center}
\end{figure*}

In Fig.~\ref{fig:filter}, we compare kernel functions $U(\theta)$ and $Q(\theta)$. As shown in Eq.~\eqref{eq:u_comp}, the kernel function $U(\theta)$ is a compensated filter, which is important to mitigate the contamination due to the LSS, except for the case of the Gaussian filter. The peak of $U(\theta)$ is located at typically a few arcminute, which corresponds to the typical angular size of massive clusters of galaxies on the sky. The corresponding kernel functions $Q(\theta)$ indicate that the tangential shear at $\theta\sim 3'-10'$ typically contributes to the signal.

In Fig.~\ref{fig:maps}, we show aperture mass maps for three different filters, Gaussian, PEX1, and TI20. It is found that the mass map for the Gaussian filter retains both large- and small-scale structures. In contrast, in mass maps with PEX1 and TI20, the large-scale structure is removed thanks to the compensated filter. For the case of TI20, the small-scale structure is also removed due to the large inner boundary of $Q(\theta)$. The TI20 setup is used to minimize various systematic effects that are important near centers of massive clusters\cite{2021PASJ...73..817O}.

The selection efficiency and completeness of clusters from aperture mass maps can be improved by combining results from several different mass maps constructed with different setups. For instance, we can combine mass maps with different source redshifts\cite{2005ApJ...624...59H,2021PASJ...73..817O,2020PASJ...72...78H} or different kernel sizes\cite{2023A&A...678A.125L}. By fully utilizing source redshift information, it is in principle possible to reconstruct three-dimensional mass maps to select clusters from their peaks\cite{2015MNRAS.449.1146L,2021ApJ...916...67L,2023arXiv231200309Y}.

\section{Cluster samples}

The first discoveries of clusters from weak lensing mass map peaks were reported in the early 2000s\cite{2001ApJ...557L..89W,2002ApJ...580L..97M}. Since then, weak lensing shear-selected cluster samples have been constructed in various surveys\cite{2005A&A...442...43H,2006ApJ...643..128W,2007A&A...462..459G,2007ApJ...669..714M,2007A&A...462..875S,2009PASJ...61..833H,2012ApJ...748...56S,2012ApJ...750..168K,2014MNRAS.442.2534S,2015PhRvD..91f3507L,2015ApJ...807...22M,2018PASJ...70S..27M}. Table~\ref{tab:samples} summarizes the history of the construction of cluster samples from weak lensing mass map peaks, focusing on clusters with high signal-to-noise ratios $\nu\geq 5$. It is found that the cluster samples have been quite small until very recently, which is explained by the fact that the construction of the large sample of shear-selected cluster samples require deep {\it and} wide imaging survey observations. On one hand, the number density of shear-selected clusters on the sky is quite sensitive to the source number density used for the weak lensing analysis (see e.g., Appendix~1 of Miyazaki {\it et al.}\cite{2018PASJ...70S..27M}), while on the other hand clusters are rare objects.

\begin{table*}[t]
  \caption{Compilation of weak lensing shear-selected cluster samples.}
  \begin{center}
    \begin{tabular}{ lccccc } \hline\hline
      Author & Data & Area & Number of clusters ($\nu\geq 5$)\\\hline
      Miyazaki {\it et al.} (2002)\cite{2002ApJ...580L..97M} & Subaru/Suprime-Cam\cite{2002PASJ...54..833M} & 2.1~deg$^2$ & $\sim 5$ \\
      Wittman {\it et al.} (2006)\cite{2006ApJ...643..128W} & DLS\cite{2002SPIE.4836...73W} & 8.6~deg$^2$ & $\sim 2$ \\
      Gavazzi \& Soucail (2007)\cite{2007A&A...462..459G} &  CFHTLS Deep\cite{2006A&A...457..841I} & 4~deg$^2$ & $\sim 2$  \\
      Miyazaki {\it et al.} (2007)\cite{2007ApJ...669..714M} & Subaru/Suprime-Cam\cite{2002PASJ...54..833M} & 16.7~deg$^2$  & 12  \\
      Shan {\it et al.} (2012)\cite{2012ApJ...748...56S} & CS82\cite{2014RMxAC..44..202M} & 64~deg$^2$  & $\sim 17$  \\
      Liu {\it et al.} (2015)\cite{2015PhRvD..91f3507L} & CFHTLenS\cite{2012MNRAS.427..146H} & 130~deg$^2$ & $\sim 10$ \\
      Miyazaki {\it et al.} (2018)\cite{2018PASJ...70S..27M} & HSC-SSP\cite{2018PASJ...70S...4A} & 160~deg$^2$ & 47 \\
      Hamana {\it et al.} (2020)\cite{2020PASJ...72...78H} & HSC-SSP\cite{2018PASJ...70S...4A} & 120~deg$^2$ & 124 \\
      Oguri {\it et al.} (2021)\cite{2021PASJ...73..817O} & HSC-SSP\cite{2018PASJ...70S...4A} & 510~deg$^2$ & 325 \\
      \hline
 \end{tabular}
\end{center}
\label{tab:samples}
\end{table*}

The breakthrough has been made by the Hyper Suprime-Cam\cite{2018PASJ...70S...1M} mounted on the Subaru 8.2-meter telescope\cite{2021PJAB...97..337I}. The Hyper Suprime-Cam Subaru Strategic Program (HSC-SSP)\cite{2021PASJ...73..817O} is a deep multi-band imaging survey covering more than 1000~deg$^2$ at the completion of the survey. It successfully finds $\mathcal{O}(100)$ shear-selected clusters from the first-year HSC-SSP data covering $\sim 160$~deg$^2$\cite{2018PASJ...70S..27M,2020PASJ...72...78H}. From the HSC-SSP Year 3 data covering $\sim 510$~deg$^2$, more than 300 shear-selected clusters are found\cite{2021PASJ...73..817O}, which represent the largest sample of shear-selected clusters constructed to date. 

\begin{figure*}[t]
\begin{center}
\includegraphics[width=0.7\hsize]{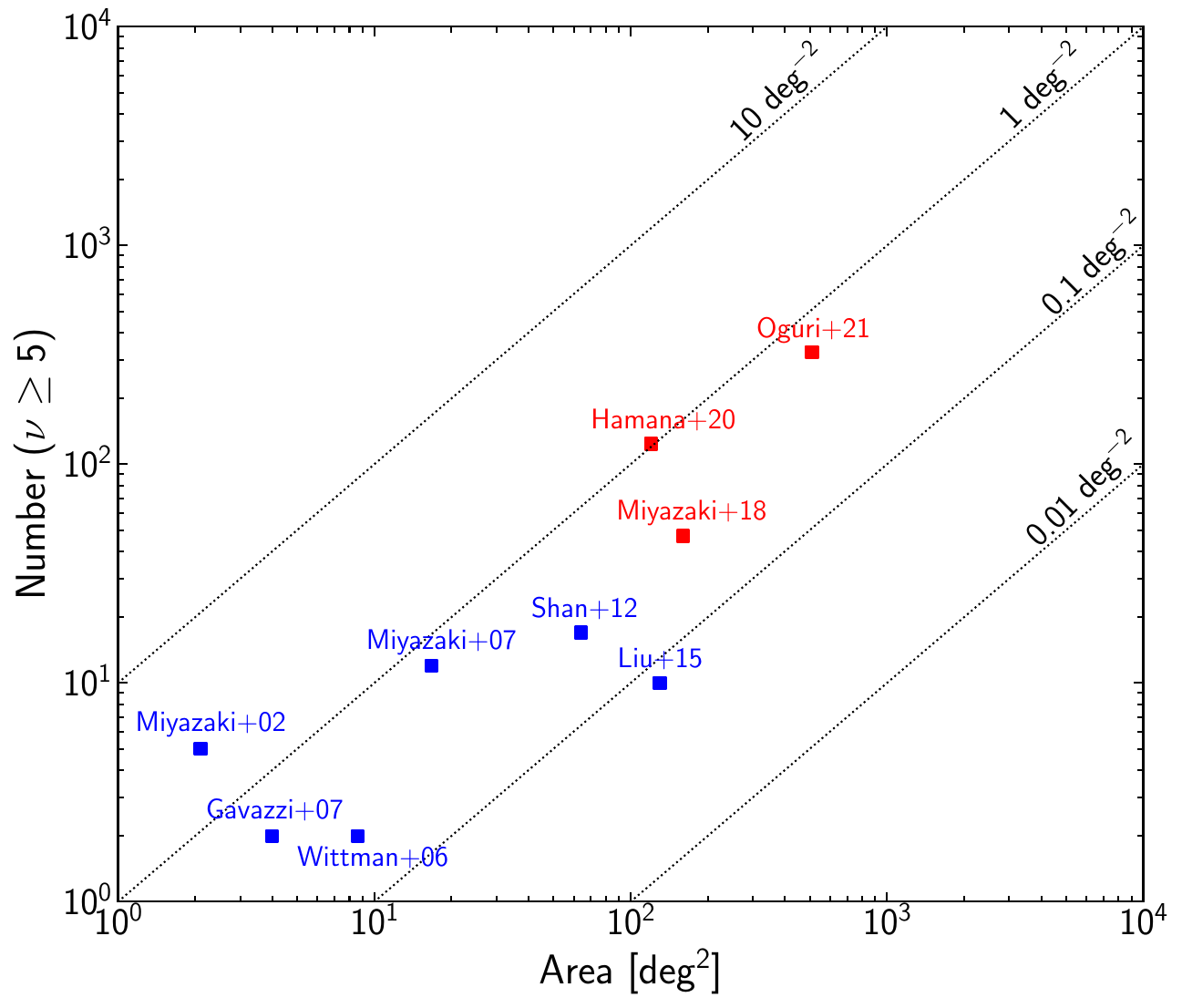}
\caption{The history of shear-selected cluster samples. See Table~\ref{tab:samples} for more details of individual samples.
\label{fig:wl_cluster_sample}}
\end{center}
\end{figure*}

The history of shear-selected cluster samples is also summarized in Fig.~\ref{fig:wl_cluster_sample}. We note that all these samples are constructed from deep optical imaging surveys with ground-based telescopes. Given the typical number density of source galaxies of $\sim 20-30$~arcmin$^{-2}$ for deep imaging with ground-based telescopes, the typical number density of shear-selected clusters on the sky from such imaging surveys is $\sim 0.3-1$~deg$^{-2}$. Thus the larger sample of shear-selected clusters can be constructed by increasing areas for such deep imaging.

\section{Applications to cluster astrophysics and cosmology}

\subsection{Baryonic properties of shear-selected clusters}

An interesting application of shear-selected cluster samples is to study baryonic properties of clusters of galaxies. Traditionally clusters of galaxies are selected based on their baryonic properties, including cluster member galaxies and the intracluster medium (ICM). A concern is that such selection based on baryonic features may miss some fraction of clusters with peculiar baryonic properties. For instance, it is known that X-ray luminosities are significantly affected by cluster mergers\cite{2002ApJ...577..579R}, suggesting that X-ray selected clusters are biased in terms of their dynamical states\cite{2011A&A...526A..79E}. Therefore, if there is a large population of X-ray underluminous clusters, a large fraction of those clusters are missed in traditional cluster searchers with X-ray observations. This has a considerable impact on using the abundance of clusters to measure cosmological parameters, and hence is important. Since shear-selected clusters do not rely on such baryonic features in constructing a cluster sample and the selection function is relatively straightforward to understand\cite{2004MNRAS.350..893H,2002ApJ...575..640W,2007A&A...471..731P}, they allow us to test whether such population of X-ray underluminous clusters exist or not\cite{2014ApJ...786..125S}.

From a systematic X-ray study of shear-selected clusters from the Subaru Suprime-Cam weak lensing survey, it is claimed that shear-selected clusters appear to be underluminous in X-ray by a factor of 2 or so compared with the expected X-ray luminosities from the known scaling relation between cluster masses and X-ray luminosities\cite{2015MNRAS.447.3044G}. A caveat is that the so-called Eddington bias is important when interpreting X-ray luminosities of shear-selected clusters\cite{2020ApJ...891..139C,2023PASJ...75...14H}. Due to the shape and LSS noises, some of low-mass clusters are up-scattered and detected as weak lensing mass map peaks. By correcting for the Eddington bias, it is argued that observed X-ray luminosities of Subaru Suprime-Cam shear-selected clusters are consistent with the expectation from the scaling relation\cite{2020ApJ...891..139C}.

The much more complete analysis of X-ray properties of shear-selected clusters is conducted by comparing shear-selected clusters from the HSC-SSP with X-ray data from the eROSITA Final Equatorial-Depth Survey\cite{2022A&A...661A..14R}. By comparing weak lensing masses and X-ray luminosities of all 25 shear-selected clusters in an overlapping sky region covering $\sim 90$~deg$^2$, it is found that the normalization of the scaling relation is consistent between shear-selected and X-ray selected clusters, as long as the selection bias is properly corrected for. This result indicates that there is no significant population of X-ray underluminous clusters, which is a good news for using X-ray selected cluster samples as an accurate cosmological probe.

In addition to X-ray properties, it is of great interest to carefully examine optical properties and the Sunyaev-Zel'dovich effect properties for a large sample of shear-selected clusters.

\subsection{Cosmology with shear-selected clusters}

The abundance of clusters of galaxies is known as a sensitive probe of cosmology\cite{2011ARA&A..49..409A}, yet challenges lie in how to measure masses of clusters and how to model the selection function. Cosmology with shear-selected clusters potentially overcome these challenges, thanks to the selection based purely on their gravitational effects including those of dark matter. Bearing this advantage in mind, theoretical studies have been conducted to check their constraining power for cosmology\cite{2009ApJ...691..547W,2009ApJ...698L..33M,2010PhRvD..81d3519K,2010MNRAS.402.1049D,2010ApJ...712..992B,2011PhRvD..84d3529Y,2012MNRAS.423..983P,2013MNRAS.432.1338M,2019ApJ...884..164Y,2022MNRAS.513.4729D,2023MNRAS.520.6382L} and to develop analytic models to predict the abundance of shear-selected clusters for each cosmological model\cite{2010A&A...519A..23M,2010ApJ...719.1408F,2015A&A...576A..24L,2015A&A...583A..70L,2016A&A...593A..88L,2016PhRvD..94h3506Z,2018ApJ...857..112Y,2018MNRAS.478.2987W}. It is found that weak lensing signals of high $\nu$ peaks are indeed dominated by single massive halos corresponding to clusters of galaxies. However, the contribution of the LSS projected along the line-of-sight is also found to have a non-negligible impact on the prediction on the abundance of weak lensing mass map peaks at high $\nu$, which needs to be taken into account for cosmological analyses. An alternative approach is to predict the abundance of mass map peaks directly from ray-tracing simulations, for which the association of mass map peaks with clusters of galaxies is not necessarily needed.

Because of the small number of shear-selected clusters as summarized in Table~\ref{tab:samples}, cosmological constraints from the abundance of shear-selected clusters have not been very competitive until recently\cite{2015PhRvD..91f3507L,2015MNRAS.450.2888L,2015PASJ...67...34H,2018MNRAS.474.1116S}. With the significant advance achieved by HSC-SSP, it is now possible to obtain tight constraint on cosmological parameters from the abundance of shear-selected clusters\cite{2023MNRAS.519..594L}.

In order to obtain accurate cosmological constraints, however, one has to take proper account of various systematic effects. One of such effects come from the construction of weak lensing mass maps. Real galaxy catalogs used for weak lensing are quite inhomogeneous due to the variation of observing conditions, and many regions such as regions around bright stars are masked. The boundary and mask of the survey region, as well as the pixelization and flat sky projection, affect weak lensing signals in a complicated manner\cite{2014ApJ...786...93U,2014ApJ...784...31L,2018A&A...614A..36L,2020A&A...638A.141P,2023A&A...671A..17A}, and these effects need to be quantified for accurate cosmological analyses. Another important effect comes from source galaxy samples, such as intrinsic alignments, cluster member dilution effects, and source clustering effects\cite{2021PASJ...73..817O,2011ApJ...735..119S,2020PASJ...72...78H,2016MNRAS.463.3653K}. It is not obvious how to model these effects in theoretical predictions of the peak abundance, and hence they are potentially an important source of systematic effects in cosmological analyses. Furthermore, weak lensing signals near centers of clusters of galaxies are subject to various systematic effects such as the non-linearity of shear\cite{2018PASJ...70S..27M} and the modification of central density profiles of clusters due to baryonic effects\cite{2013PhRvD..87b3511Y,2015ApJ...806..186O,2019MNRAS.488.3340F,2020MNRAS.495.2531C,2021MNRAS.502.5593O}.

Recently Chen {\it et al.}\cite{2024arXiv240611966C} propose a new approach to derive robust cosmological constraints from shear-selected cluster samples. To mitigate various systematic effects summarized above, this approach adopts a conservative cut of source galaxies to select galaxies at $z> 0.7$, aiming at selecting galaxies located behind most of shear-selected clusters, and chooses the kernel function that minimizes contributions from centers of clusters of galaxies (specifically TI20 filter mentioned above). In addition, to take full account of complicated uncertainties associated with the construction of weak lensing mass maps, such as the variation of observing conditions and masking effects and the resulting inhomogeneity of noise properties of mass maps as discussed in Sec.~\ref{sec:sn_ratio}, this approach conducts semi-analytical injection simulations for which NFW halos are injected into real weak lensing mass maps. Taking advantage of this new approach, Chiu {\it et al.}\cite{2024OJAp....7E..90C} derive cosmological constraints from HSC-SSP Year 3 shear-selected cluster catalogs to find $\hat{S}_8\equiv \sigma_8(\Omega_{\mathrm{m}}/0.3)^{0.25}=0.835^{+0.041}_{-0.044}$. This work demonstrates that accurate cosmological constraints can indeed be obtained from shear-selected cluster catalogs.

\subsection{Comparison with peak statistics}

More popular approach to derive cosmological constraints from weak lensing mass maps is the so-called peak statistics for which low signal-to-noise ratio peaks are mainly used to derive cosmological constraints\cite{2016MNRAS.463.3653K,2021MNRAS.506.1623H,2022MNRAS.511.2075Z,2024MNRAS.528.4513M,2024arXiv240510312H}. In the peak statistics, mass map peaks are regarded as observables that can be predicted from cosmological models e.g., using ray-tracing simulations, and therefore no association of mass map peaks with astronomical objects such as clusters of galaxies or galaxies are not assumed. We note that cosmology with shear-selected clusters have its own advantages over the peak statistics. The fact that weak lensing signals are dominated by single massive clusters indicate that we have better control of systematics arising from e.g., the member dilution effect, intrinsic alignment, and baryonic effect, by carefully choosing the set-up to construct aperture mass maps. Theoretical modeling of the abundance of shear-selected clusters is more straightforward. Given the increasing importance of controlling various systematic effects in cosmological analyses, it is of great interest to carefully compare results from low $\nu$ and high $\nu$ peaks in weak lensing mass maps.

\section{Conclusion}

Peaks in weak lensing aperture mass maps provide a new route to identify clusters of galaxies. This selection is based on gravitational lensing effects of the total mass distribution including dark matter, and is highly complementary to traditional approaches to select clusters of galaxies from member galaxies or ICM properties. Such shear-selected cluster samples have several important applications, including studies of baryonic properties of clusters and cosmology.

The applications of shear-selected clusters have been hampered by the small number of such clusters. This is because of high requirements for observations. In order to construct a large sample of shear-selected clusters, it is essential to conduct deep {\it and} wide imaging surveys, because the number density of shear-selected clusters is quite sensitive to the source number density of galaxies used for weak lensing analyses, and also because clusters are rare objects on the sky.

Recently, the significant advance is achieved by the HSC-SSP, in which more than 300 of shear-selected clusters with the signal-to-noise ratio $\nu\geq 5$ are already discovered. It is shown that accurate cosmological constraints can be derived from such cluster samples, by carefully choosing the kernel functions as well as background source galaxy samples to mitigate various systematic effects.

In the next decade, we can expect rapid progress in increasing the number of shear-selected clusters, thanks to massive imaging surveys conducted by {\it Euclid}\cite{2024arXiv240513491E}, the {\it Rubin Observatory} Legacy Survey of Space and Time\cite{2019ApJ...873..111I}, and the {\it Nancy Grace Roman Space Telescope}\cite{2021MNRAS.507.1746E}. 

\section*{Acknowledgments}
We thank anonymous referees for careful reading of the manuscript and for constructive comments.
This work was supported by JSPS KAKENHI Grant Numbers JP20H05856, JP22K21349, JP19KK0076, JP23K22531, JP24K00684.



\end{document}